**Nonlinear van der Waals metasurfaces with resonantly enhanced light generation**

*Haonan Ling[1], Yuankai Tang[2], Xinyu Tian[2], Pavel Shafirin[1], Mozakkar Hossain[1], Polina P. Vabishchevich[3], Hayk Harutyunyan[2*] & Artur R. Davoyan[1*]*

1. Mechanical and Aerospace Engineering Department, University of California, Los Angeles, CA, 90095, USA
2. Department of Physics, Emory University, GA, 30322, USA
3. Department of Chemistry and Biochemistry, University of Maryland, College Park, MD 20742, USA

*[*]Correspondence to: hayk.harutyunyan@emory.edu and davoyan@seas.ucla.edu*

**Abstract:** Efficient nonlinear wave mixing is of paramount importance for a wide range of applications. However, weak optical nonlinearities pose significant challenges for accessing nonlinear light-matter interaction in compact systems. Here, we experimentally study second harmonic generation in deeply subwavelength 3R-MoS$_2$ metasurfaces ($< \lambda/13$ thick). Our measurements, supported by theoretical analysis, reveal a complex interplay and coupling between geometric resonances, optical extinction, and strong nonlinear susceptibility dispersion near excitons. We further demonstrate $> 150$-fold enhancement in second harmonic signal at $740\ nm$ driven by the A exciton resonance. Additionally, our theoretical studies predict an enhancement of more than $10^6$ in second harmonic generation in <100 nm thick structures exhibiting bound states in the continuum resonance. These findings provide insight into accessing and harnessing the unprecedented 3R-MoS$_2$ nonlinearities at a subwavelength scale, paving the way for ultracompact nonlinear photonic devices.

**Main text**

Nonlinear wave mixing is of great importance for quantum [1-4] and classical [5-11] light generation, and for amplification and detection of weak optical signals [12-16]. Due to its versatility, nonlinear optics plays a key role in a wide range of fields, from telecommunications [17-19], to spectroscopy [20,21] and microscopy [22-24], and to information processing and all-optical computing [25,26]. However, achieving efficient three-wave mixing, particularly second harmonic generation (SHG), is hindered by weak optical nonlinearities and the requirement for phase matching between interacting waves [27]. As a result, SHG systems often rely on bulky structures that are difficult to scale and integrate with advanced photonic architectures [6,28-30]. Resonant metasurfaces offer an alternative approach to enhancing light-matter interactions in a compact format [31-34]. By engineering optical resonances and mode symmetry, metasurfaces serve as a functional tool for nonlinear light generation and mixing without the need for complex phase matching techniques [31,34,35]. A broad variety of nonlinear phenomena have been successfully demonstrated, including second and third harmonic generation [32,33,36-41], parametric down-conversion [4,42,43], and high-harmonic generation [44-46]. Recent demonstrations of ultrahigh quality factor metasurfaces [47-49], governed by the bound states in the continuum (BIC) resonances, further enable drastic enhancement of nonlinear interactions and harmonic signal generation [50-53].

Nonetheless, even in well-optimized structures, the efficiency of nonlinear generation is inherently constrained by the intrinsic properties of the underlying materials [27]. In SHG, the strength of the nonlinear interaction is determined by the second-order susceptibility tensor, $\bar{\bar{\chi}}^{(2)}$. Since the second harmonic power scales as $\left|\bar{\bar{\chi}}^{(2)}\right|^2$, selecting materials with high $\bar{\bar{\chi}}^{(2)}$ tensor elements can significantly enhance nonlinear interaction and improve generation efficiency. LiNbO$_3$, a conventional material for accessing second-order effects, supports moderate $\bar{\bar{\chi}}^{(2)}$ tensor values of approximately $50\ pm/V$ in the near-infrared frequency range (with $d_{33} = -25.2\ pm/V$) [54]. Combined with its relatively low refractive index ($n \simeq 2.1$), systems based on LiNbO$_3$ require long interaction lengths to achieve high nonlinear conversion efficiencies [29,30]. Metasurfaces made of LiNbO$_3$ feature quality factors limited by the small refractive index contrast, which restricts SHG efficiency [38,55-57]. III-V semiconductors, such as GaAs, have higher refractive index and exhibit much higher second order susceptibilities ($\sim 250\ pm/V$ below bandgap at $\sim 1\ \mu m$ and exceeding 800 pm/V at absorption resonance deep within the bandgap for GaAs [54,58,59]). SHG in GaAs and GaP metasurfaces has been successfully demonstrated and explored in several prior studies [32,60-62]. However, due to the $\bar{4}3m$ symmetry of GaAs, nonlinear generation involves complex $x - y - z$ polarization mixing of the interacting fields. Additionally, metasurfaces made of conventional covalent 3D materials require designs with high refractive index contrast to efficiently localize optical modes, which often necessitates complex growth and fabrication [63].

Featuring smooth, self-passivated surfaces and free from lattice-matching constraints, layered van der Waals transition metal dichalcogenides (TMDCs) offer a promising alternative [64-66]. Their high refractive index (figure 1b) and emergent optical and electronic properties make TMDCs an appealing semiconductor platform for designing functional metasurfaces [67-70] and subwavelength integrated photonic devices [71-74]. Monolayer TMDC films exhibit inversion symmetry breaking, which, combined with sharp exciton resonances, leads to large second order nonlinear susceptibility [75-77]. $\chi^{(2)}$ values exceeding 1000 pm/V, when excited close to the exciton resonance, have been reported for select monolayer TMDC compounds (particularly MoSe$_2$, MoTe$_2$, and WS$_2$) [77-79]. However, due to the naturally small cross-section of light-matter interaction in monolayer films, the overall efficiency of nonlinear light conversion is limited. It has been suggested that heterostructure engineering could maintain the inversion symmetry breaking while increasing the nonlinear interaction volume [80-82]. More recently, it was discovered that the rhombohedral (3R) phase of molybdenum disulfide (MoS$_2$), due to its asymmetric stacking, preserves the broken inversion symmetry even in bulk crystals (figure 1a, inset) [83,84]. Consequently, several studies have examined SHG in thin 3R-MoS$_2$ films and planar waveguides, where geometrically induced phase-matching conditions were explored [83-85]. Spectroscopic characterizations of thin films with thickness less than 100 nm reveal strong frequency dependence of second harmonic intensity, which peaks near $A$ and $B$ exciton resonances [83,85]. Additionally, a significantly stronger SHG was observed at $\lambda_{SHG} \simeq 455\ nm$ (i.e., $\lambda_{pump} \simeq 910\ nm$) in planar thin films [84-86]. Subsequent density functional theory calculations of $\bar{\bar{\chi}}^{(2)}$ tensor confirmed its extremely large dispersion, showing that the principal tensor components reach $\simeq 800\ pm/V$, making 3R-MoS$_2$ one of the strongest nonlinear materials [86].

The use of resonant nanostructures could further enhance SHG [70,86,87]. Studies of SHG in individual nanoresonators made of 3R-MoS$_2$ featuring high quality anapole resonances [86] have demonstrated a more than 100-fold enhancement of the generated signal compared to planar unpatterned 3R-MoS$_2$ films. While individual resonators allow exploring intricate aspects of nonlinear light-matter interaction, their overall interaction cross-section with the incoming laser beam is inherently small. Additionally, pumping around and above MoS$_2$ bandgap (i.e., $\lambda_{pump} < 1000\ nm$) may lead to undesired parasitic absorption, which could limit nonlinear interaction. Understanding and harnessing below-bandgap resonant light-matter interaction for nonlinear signal generation in 3R-MoS$_2$ remains highly unexplored.

Here, we demonstrate the resonant interplay and enhancement of SHG in deeply subwavelength 3R-MoS$_2$ metasurfaces (thickness $\leq \lambda_{pump}/13$). By tuning metasurface mode dispersion and probing second harmonic emission at the $A$ and $B$ excitons, as well as further into the bandgap, we reveal a complex three-fold coupling between geometric resonances, linear extinction, and exciton-assisted resonances in $\bar{\bar{\chi}}^{(2)}$. We observe an over 150-fold enhancement in SHG at $\lambda_{SHG} \simeq 740\ nm$ compared to a planar film of the same thickness ($\lambda_{pump}/13$). Our theoretical studies further predict that more than 6 orders of magnitude enhancement in SHG is achievable by exploiting BIC resonances. This work provides insights on accessing and harnessing unprecedented MoS$_2$ nonlinearities at the subwavelength scale.

Figure 1a shows a schematic of our experiment. In this setup, a 3R-MoS$_2$ metasurface, placed on a fused silica substrate, is excited from above by a 140-fs laser beam produced by an optical parametric oscillator (OPO) (with an output range of 1000 nm – 1600 nm; details provided in the Methods section). This setup allows for pumping below the 3R-MoS$_2$ bandgap (i.e., $\lambda_{pump} > 1000\ nm$) and probing second order light-matter interactions above the bandgap, including at $A$ and $B$ excitons and extending deeper into the bandgap (i.e., in $500\ nm < \lambda_{SHG} < 800\ nm$ range; see also figure 1b).

We anticipate that pumping at the metasurface resonances, $\lambda_R$, could lead to enhanced second harmonic generation (i.e., when $\lambda_{pump} \simeq \lambda_R$ resonant SHG at $\lambda_{SHG_R} \simeq \lambda_R/2$ is expected). Therefore, by controlling the metasurface resonances, $\lambda_R$, resonant SHG can be tuned to cross both of 3R-MoS$_2$'s excitons (at $\lambda_B \approx 625\ nm$ and $\lambda_A \approx 675\ nm$ [83,85]; see figure 1b). To achieve this, we study metasurfaces with two different periods: $0.85\ \mu m$ and $1\ \mu m$. In figure 1c we plot the calculated electric field intensity inside the disk volume (i.e., $\int |\boldsymbol{E}_\omega|^2 dV$, where $\boldsymbol{E}_\omega$ is the pump electric field) as a function of the incident laser wavelength and disk radius for a $0.85\ \mu m$ period and $110\ nm$ thick metasurface. The corresponding response for a $1\ \mu m$ period metasurface is shown in figure S6. In both cases, the metasurfaces exhibit a distinct resonance at $\lambda > 1100\ nm$. This resonance corresponds to the excitation of a dipolar Mie mode with the electric field predominantly polarized along the 3R-MoS$_2$ crystallographic plane (figure 1d, inset). For a smaller period structure ($0.85\ \mu m$), the fundamental metasurface resonance can be tuned to spectrally align the expected SHG enhancement with the $A$ and $B$ excitons (i.e., $\lambda_{SHG_R} \simeq \lambda_{A,B}$). A complex interplay between nonlinear generation and exciton resonances is anticipated in this case. In contrast, the generated second harmonic wavelength for the larger period structure (1 $\mu m$) falls below the excitons (i.e., $\lambda_{SHG_R} > \lambda_A$) for any disk radius when pumped at the metasurface resonance (figure S6). As a result, a significantly different interaction scenario is anticipated.

Apart from the stand-alone branch at $\lambda > 1100\ nm$, several other features are worth noting for both 0.85 $\mu m$ and 1 $\mu m$ period structures at shorter wavelengths. In the ~700 $nm < \lambda <$ ~1100 $nm$ range, where optical absorption is relatively small [85,88], several resonances, traceable to a combination of Fabry-Perot and Mie excitations, are observed [89]. Notably, the shorter wavelength mode, excited just above the $A$ exciton, supports a doubly resonant interaction at both the fundamental frequency and the corresponding second harmonic (figure 1c and figure S6). At even shorter wavelengths, starting with the $A$ exciton and extending deeper into the bandgap (i.e., for $\lambda < 700\ nm$), strong optical absorption precludes any noticeable excitations in the structure.

Guided by numerical simulations, we fabricate metasurface samples with three selected disk radii for each period. Figure 1a (inset) and figure S4 show scanning electron micrographs of the fabricated metasurfaces (fabrication details are discussed in Methods and SI section II). Figure 1d presents the measured transmission spectra for the 0.85 $\mu m$ period metasurface (corresponding measurements for the 1 $\mu m$ period metasurfaces are shown in figure S8). As designed, all samples exhibit several resonances in the 700 nm – 1000 nm range, with a distinct stand-alone resonance at $\lambda > 1300\ nm$. The measured resonance positions for the 0.85 $\mu m$ metasurface are ~1363 nm, ~1488 nm, and ~1579 nm for fabricated structures with ~0.267 $\mu m$, ~0.305 $\mu m$, and ~0.355 $\mu m$ disk radii, respectively (also marked on the colormap in figure 1c). The measured spectra agree well with corresponding numerical simulations (see also figure S7). Minor discrepancies between theory and experiment can be attributed to uncertainties in materials data [85] and fabrication imperfections.

Next, we examine SHG in the fabricated metasurface samples. A $\approx 0.8\ mW$ beam from an OPO is focused into a ~2 $\mu m$ diameter spot (see also Methods for more details). We then scan pump wavelength from 1000 nm to 1600 nm with a 10 nm increment and collect the generated second harmonic signal in reflection (figure 1a). The corresponding spectrally integrated second harmonic signal for metasurfaces with 0.85 $\mu m$ and 1 $\mu m$ period are plotted in figures 2a and 2b, respectively. The second harmonic nature of the generated signal is verified by a standard power law fitting (see SI section I and figure S2 for details). For comparison, SHG from a 110 nm thick unpatterned $MoS_2$ film is also plotted in both figures 2a and 2b (see SI section V for details on SHG from unpatterned $MoS_2$ thin films).

Several characteristic features in the generated second harmonic signal can be highlighted. First, similar to a planar 3R-$MoS_2$ film, all six of the measured samples (shown in figures 2a and 2b) exhibit second harmonic intensity peaks around $\lambda_{SHG} \approx 580\ nm$, $\approx 620\ nm$, $\approx 670\ nm$. The 620 $nm$ and 670 $nm$ peaks correlate well with the positions of $B$ and $A$ excitons observed in the linear extinction spectra (figure 1b). The origin of the 580 nm resonance can be attributed to the dispersion of the susceptibility tensor itself [86], which has also been observed in unpatterned 3R-MoS$_2$ thin films in previous studies [84-86]. The strength of the second harmonic signals at these three wavelengths varies significantly with the structure dimensions. Additionally, at $\approx 670\ nm$, the SHG is significantly enhanced compared to a planar film of the same thickness (by more than 30 times for $0.85\ \mu m$ and over 10 times for $1\ \mu m$ period structures).

Second, for all measured samples, we observe a peak in the SHG signal at a wavelength close to half of the metasurface resonance wavelength, i.e., at $\lambda_{SHG} \simeq \lambda_R/2$ (the corresponding positions of $\lambda_R/2$ are marked by arrows in figures 2a and 2b, respectively). A slight shift of the SHG peak position relative to $\lambda_R/2$ can be attributed to the strong dispersion of the linear refractive index (around $\lambda_R/2$), especially near the A exciton, as well as the dispersion of the nonlinear susceptibility at the corresponding fundamental frequency ($\lambda_{pump} \simeq 2\lambda_A$). This metasurface-induced resonant light-matter interaction leads to a significant enhancement of the generated second harmonic signal. For example, the $0.85\ \mu m$ period metasurface with a $0.305\ \mu m$ disk radius exhibits more than 150-fold increase in second harmonic signal at $\lambda_{SHG} \simeq 740\ nm$ compared to a planar film.

Third, for all measured samples, we observe several additional SHG peaks at $\lambda_{SHG} \geq 690\ nm$, which do not have clear analogues in the planar film and cannot be directly matched with metasurface resonances. We attribute these excitations in part to the dispersion of $\bar{\bar{\chi}}^{(2)}$ tensor components, which due to symmetry protection, are not readily accessible in simple planar films. To better understand the complex interplay at work, we perform numerical simulations that account for the full tensorial nature of the second-order susceptibility tensor.

In our simulations we use data for the $\bar{\bar{\chi}}^{(2)}$ tensor previously obtained by the density functional theory [86]. We assume that the metasurface is excited by a plane continuous wave incident from the top. By sweeping the excitation wavelength, $\lambda_{pump}$, we examine generation at the second harmonic (see Methods). Figures 2c and 2d show the calculated spectrally resolved second harmonic intensity in logarithmic scale as a function of disk radius for $0.85\ \mu m$ and $1\ \mu m$ period metasurfaces, respectively. For convenience, we also mark peaks of measured experimental second harmonic spectra in figures 2c and 2d. The numerical simulations agree well with the observed experimental data in all measured cases (discrepancies can be attributed in part to light collection losses in our measurements). For both metasurface periods, a distinct resonance branch at a longer wavelengths ($\lambda_{SHG} > 620\ nm$ for $0.85\ \mu m$ period and $\lambda_{SHG} > 710\ nm$ for $1\ \mu m$ period), originating from the metasurface resonance at the fundamental frequency, is clearly visible (see also figure 1c and figure S6). As designed, for the $0.85\ \mu m$ period this branch crosses A and B excitons, and a corresponding weak coupling is observed (figure 2c). Exciton-assisted resonances are more pronounced for the $1\ \mu m$ period metasurface (figure 2d), where a complex evolution involving resonance splitting and crossing as a function of disk radius is evident. As the disk radius changes, so do the profile and strength of the excited field at both fundamental and second harmonic frequencies. This results in an intricate dynamic in SHG, driven by the convolution of $\bar{\bar{\chi}}^{(2)}$ tensor dispersion [86] and evolving mode profiles. The resonance splitting near the *A* and *B* excitons for the 1 $\mu m$ period metasurface is also clearly observed in our measurements.

To better understand the interplay between the *A* and *B* excitons and the related nonlinear susceptibility dispersion, we numerically examine SHG as a function of metasurface thickness, assuming 0.85 µm period structure with a 0.3 µm disk radius. The calculated SHG spectra are shown in figure 3a. As expected, a single distinct resonance in SHG is observed for all metasurface thicknesses at these dimensions (see also figures 2a and 2c; note a linear scale used in figure 3a, compared to a log scale in figure 2c). We also observe a clear red-shift in the SHG peak as the thickness increases. This shift is attributed to the red-shift of the metasurface resonance ($\lambda_R$), resulting from the larger effective meta-atom volume in thicker structures. At the same time, the intensity of the generated second harmonic varies non-monotonically with metasurface thickness. We attribute this behavior to the combined effect of three factors: increased extinction at *A* and *B* excitons, a peak in nonlinear susceptibility at $\lambda_{pump} \simeq 1315\ nm$ [86], and thickness-dependent variations in the field strength at the fundamental frequency. The corresponding linear extinction spectra ($\kappa(\lambda_{SHG})$) and nonlinear susceptibility of bulk 3R-MoS$_2$ ($\chi^{(2)}_{xxy}(2\lambda_{SHG})$ component) are also plotted in figure 3a. The strong SHG around $\simeq 650\ nm$, observed for the 80 $nm$ thick structure, correlates well with the extinction minimum and second order susceptibility maximum. For larger thicknesses (>110 nm), the SHG peak shifts to $\lambda_{SHG} > 700\ nm$, i.e., below the *A* exciton. In this wavelength range, optical extinction is low and the associated susceptibility dispersion is weak (figure 3a). As a result, in this range, the variation of the SHG signal strength is primarily due to the excited field, which changes significantly with the structure's dimensions [85].

Finally, we explore the potential use of the strong light-matter interaction observed around $\lambda_{SHG} \simeq 650\ nm$ in figure 3a. For a conceptual demonstration, we design a high-quality factor metasurface with a BIC resonance (figure S13). Our design follows previously reported strategies [61,90]. Supported by a high-refractive index of 3R-MoS$_2$ (figure 1b), even a 90 $nm$ thick metasurface features a quasi-BIC resonance with a quality factor exceeding 1100 at $\lambda = 1313\ nm$ (see SI section VI and figure S13 for detailed structure design). Since the generated second harmonic power scales quadratically with the quality factor (i.e., $P^{2\omega} \propto (QP^{\omega})^2$ ), a dramatic enhancement in SHG is expected when pumped at the quasi-BIC resonance. Figure 3b shows the calculated second harmonic emission from a planar film and the designed quasi-BIC metasurface (both 90 $nm$ thick). A nearly six orders of magnitude stronger second harmonic signal is predicted for a quasi-BIC metasurface compared to a planar film. This combination of ultrahigh-Q metasurfaces [61,90] and the unprecedented nonlinearities of 3R-MoS$_2$ presents an opportunity for efficient nonlinear signal generation in subwavelength structures.

In our work we have investigated the intricate interaction and enhancement of SHG emission across the *A* and *B* excitons in 3R-MoS$_2$. In this wavelength range the combination of geometric resonances, exciton absorption, and strong second order susceptibility dispersion leads to a complex nonlinear interplay. The experimentally measured enhancement in SHG, compared to a planar film, exceeds 150 times at $\lambda_{SHG} \simeq 740\ nm$ in structures that are less than $\lambda/13$ thick. Even greater enhancement, over $10^6$, is predicted for high quality factor metasurfaces. We anticipate that some of our findings could extend to understanding nonlinear light-matter interactions near the $\lambda_{SHG} \simeq 455\ nm$ resonance in $\bar{\bar{\chi}}^{(2)}$ [86]. While our measurement setup cannot access this wavelength, we have probed SHG emission down to $\lambda_{SHG} = 500\ nm$. For both metasurfaces and planar films, we see a drastic increase in SHG at $\lambda_{SHG} = 500\ nm$ (see SI section V and VI for details). Even stronger signal is anticipated at $\sim 455\ nm$, and properly designed metasurfaces could further boost SHG at this wavelength. However, understanding the limits of SHG enhancement in presence of strong interband absorption will require further investigation.

**Methods**

Fabrication

The nanofabrication workflow for 3R-MoS$_2$ metasurfaces is shown in figure S3. We begin by exfoliating 3R MoS$_2$ crystals (HQ Graphene) onto a fused silica substrate (University Wafer) using the standard scotch tape method. To pattern the flakes into nanostructures, we first deposit a thin layer of gold (about 20 nm) via sputter coating (Denton Vacuum Desk V Sputter). This gold layer serves as a charge dissipation layer, an adhesion layer, and an etching mask. Next, we spin-coat a thin layer of negative e-beam resist, MaN 2403, at 3000 rpm for 30 second, resulting in a resist thickness of approximately 300 nm. The chip is then baked at 95 °C for 1 minute before exposure. E-beam lithography (Raith EBPG 5000+ES) with 100 keV beam energy is used to write patterns onto the target flake. Afterward, the flake is developed with MF-319 solution for 60 seconds, followed by a 90-second rinse in deionized water. The resist exposed by the e-beam remains on the flake after development, serving as an etching mask as well. Next, chlorine reactive ion etching (RIE) (PlasmaTherm SLR 770 ICP) is used to remove the gold layer (recipe: 24 sccm Ar, 5 sccm Cl$_2$, RF power of 100 W and ICP RF power of 250 W, etching time 15s). Fluorine RIE (Oxford PlasmaLab 80+) is then used to etch exposed 3R MoS$_2$ (recipe: 40 sccm CHF$_3$, 4 sccm O$_2$, forward power of 55 W and pressure at 40 mtorr). After the exposed MoS$_2$ is etched away, the remaining resist is removed by immersion in hot acetone, and the gold layer is washed away using Gold Etchant TFA.

Linear Transmission Measurement

A broadband white light lamp is used as the incident light source. The transmitted light is collected by an Olympus LMPlanFL N objective (NA=0.5, 50x). The signal is detected by an Andor Shamrock spectrograph with a Newton EMCCD and an iDus InGaAs detector, for visible and near-infrared spectra, respectively. A linear polarizer is placed in the collection light path to align with the periodic directions of the array in the sample.

SHG Measurement

A Coherent Chameleon Ultra II Ti:sapphire laser was used to generate 140 femtosecond (fs) pulses with a repetition rate of 80 MHz. The tunable wavelength range from 1000nm to 1600nm was achieved using a Coherent Compact optical parametric oscillator (OPO). The laser was focused on the sample at using an Olympus MPlanFL N objective (NA=0.9, 100x), resulting in a spot diameter of 1.4-2.2 um. The incident laser power was set to 0.8mW. The SHG signals were collected by the same objective and detected by an Andor Shamrock spectrograph with a Newton EMCCD. A 1000 nm long-pass filter and a 1000nm short-pass filter are placed before and after the sample, respectively, to ensure only SHG signals reached the EMCCD.

Numerical Modeling

COMSOL Multiphysics software was used for both linear and nonlinear numerical simulations. In our simulations we utilize refractive index data from Ref. [84] and the full $\bar{\bar{\chi}}^{(2)}$ tensor from Ref. [85]. Frequency domain analysis is employed in our modeling. Excitation by a plane continuous wave is assumed at the fundamental frequency, and the corresponding reflection, transmission, and field strength inside the structure are examined. For second harmonic generation, the field excited at the fundamental frequency serves as a source of nonlinear polarization density (assuming the undepleted pump approximation). The total SHG intensity is integrated over the top surface of the simulation domain. By sweeping pump wavelength, spectrally resolved second harmonic generation is obtained. To ensure comparison between different structures, the incident power is kept constant. See also SI section III for more details on simulations.

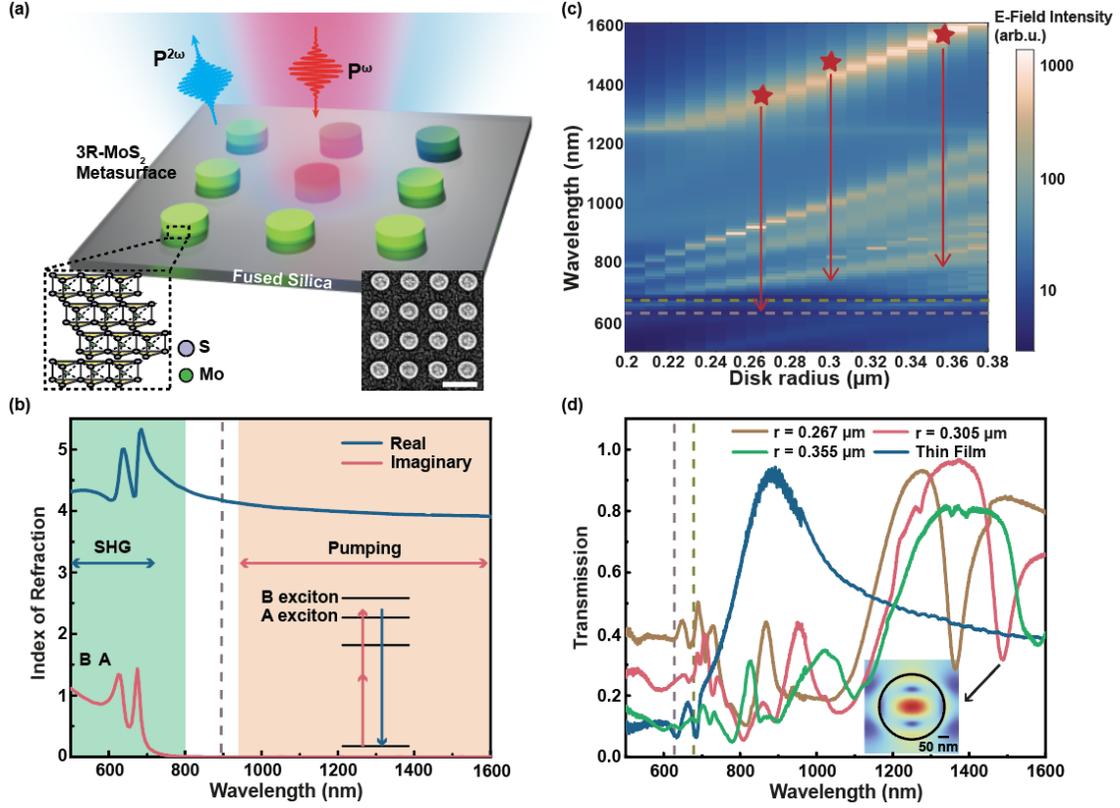

*Figure 1 | 3R-MoS$_2$ metasurfaces for probing resonant nonlinear generation.* *(a) Schematic illustration of the studied geometry. Inset shows an SEM micrograph of the fabricated sample with 0.85 μm period (Scale bar: 1 μm). (b) Optical constants of a bulk 3R-MoS$_2$ [85], showing only the in-plane components of the refractive index and extinction coefficient. The dashed line indicates the onset of the bandgap at ~960 nm for bulk 3R-MoS$_2$. In this work, we pump below the bandgap (orange area) and probe light-matter interaction at the second harmonic (green area). (c) Calculated electric field intensity, integrated over the disk volume, as a function of excitation wavelength and disk radius for a* 110 nm *thick metasurface with a* 0.85 μm *period. The two dashed lines indicate positions of A and B excitons for bulk 3R-MoS$_2$. The stars mark the positions of experimentally measured optical resonances in the fabricated samples, as shown in panel (d). (d) Measured transmission spectra for metasurfaces with disk radii of* 0.267 μm, 0.305 μm, *and* 0.355 μm *(thickness is* 110 nm *and period is* 0.85 μm*). The transmission spectrum for a planar film of the same 110 nm thickness is also shown. Inset: excited field intensity profile for a metasurface with* 0.305 μm *disk radius at* $\lambda = 1490\ nm$. *The two dashed lines indicate the positions of A and B excitons for bulk 3R-MoS$_2$.*
9

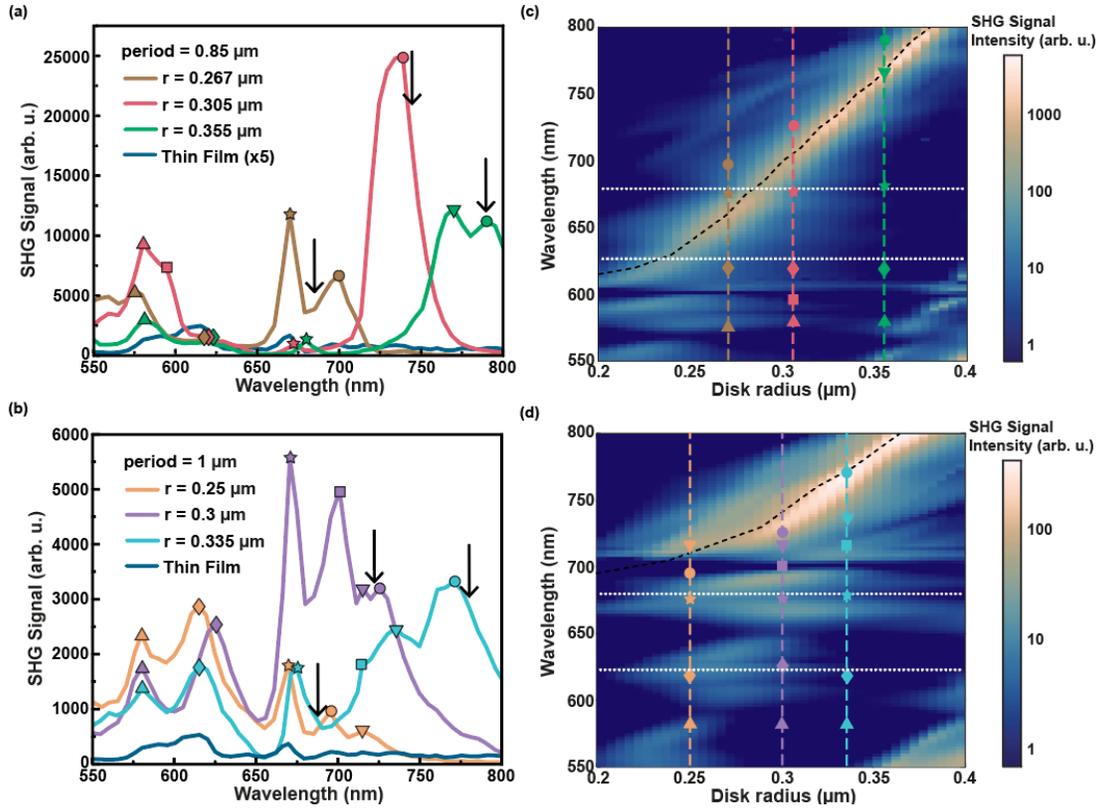

***Figure 2 | Second harmonic generation and interaction with excitons.** (**a**) Measured spectrally integrated SHG signal from a 0.85 μm period metasurface. Arrows indicate the positions of corresponding resonances at the fundamental frequency, and markers highlight peaks that are shown on a calculated colormap in (c). (**b**) Same as in panel (a), but for a 1 μm period metasurface. Markers indicate peaks highlighted on the calculated colormap in (d). (**c**) Calculated SHG for a 0.85 μm period metasurface. Markers denote the corresponding positions of the measured peaks and features. The black dotted curve represents half of the spectral position of fundamental metasurface resonances at different disk radii. (**d**) same as in panel (c) but for a 1 μm period metasurface. In both (c) and (d) white dotted lines denote the positions of excitons.*

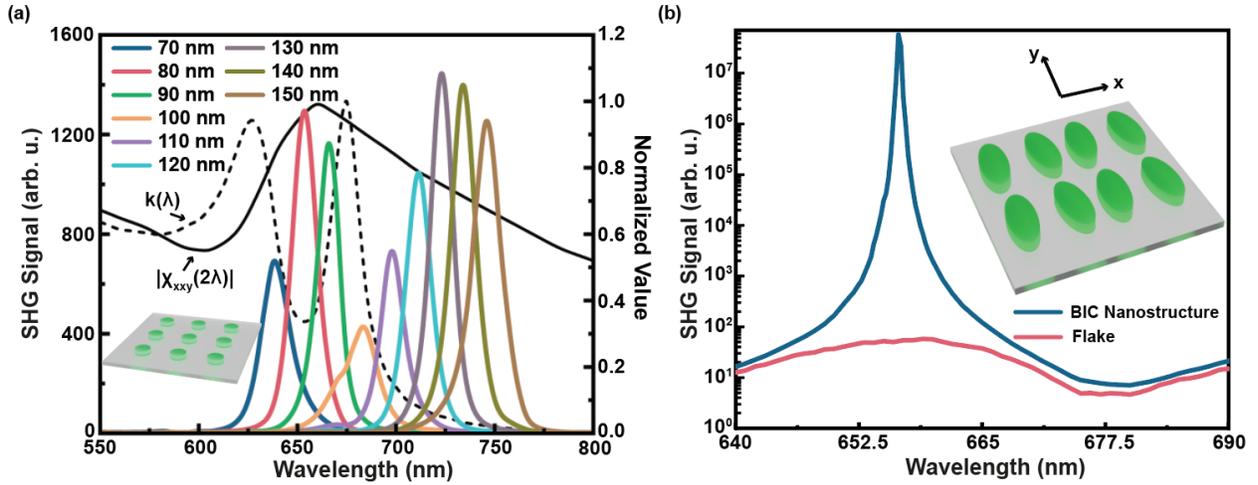

*Figure 3 | **Understanding resonant enhancement of SHG.** (a) Calculated SHG spectra for a metasurface with a 0.85 μm period and 0.3 μm disk radius as a function of metasurface thickness. The normalized 3R-MoS$_2$ extinction coefficient (in-plane component) and $\chi^{(2)}_{xxy}$ component are also plotted. (b) Calculated SHG emission from a 90 nm thick BIC metasurface compared to a planar film of the same thickness.*

# References

1. Wu, L.-A., Kimble, H., Hall, J. & Wu, H. Generation of squeezed states by parametric down conversion. *Physical Review Letters* **57,** 2520 (1986).

2. Guo, X. *et al.* Parametric down-conversion photon-pair source on a nanophotonic chip. *Light: Science & Applications* **6,** e16249–e16249 (2017).

3. Kwiat, P. G. *et al.* New high-intensity source of polarization-entangled photon pairs. *Physical Review Letters* **75,** 4337 (1995).

4. Santiago-Cruz, T. *et al.* Resonant metasurfaces for generating complex quantum states. *Science* **377,** 991–995 (2022).

5. Wang, Y., Holguín-Lerma, J. A., Vezzoli, M., Guo, Y. & Tang, H. X. Photonic-circuit-integrated titanium: sapphire laser. *Nature Photonics* **17,** 338–345 (2023).

6. Yang, J. *et al.* Titanium: sapphire-on-insulator integrated lasers and amplifiers. *Nature* **630,** 853–859 (2024).

7. Kippenberg, T. J., Holzwarth, R. & Diddams, S. A. Microresonator-based optical frequency combs. *Science* **332,** 555–559 (2011).

8. Cundiff, S. T. & Ye, J. Colloquium: Femtosecond optical frequency combs. *Reviews of Modern Physics* **75,** 325 (2003).

9. Goulielmakis, E. & Brabec, T. High harmonic generation in condensed matter. *Nature Photonics* **16,** 411–421 (2022).

10. Dudley, J. M., Genty, G. & Coen, S. Supercontinuum generation in photonic crystal fiber. *Reviews of Modern Physics* **78,** 1135–1184 (2006).

11. Wang, F., Martinson, A. B. & Harutyunyan, H. Efficient nonlinear metasurface based on nonplanar plasmonic nanocavities. *ACS Photonics* **4,** 1188–1194 (2017).

12. Schmidt, B. E. *et al.* Frequency domain optical parametric amplification. *Nature Communications* **5,** 3643 (2014).

13. Ledezma, L. *et al.* Intense optical parametric amplification in dispersion- engineered nanophotonic lithium niobate waveguides. *Optica* **9,** 303–308 (2022).

14. Pelc, J. S. *et al.* Long-wavelength-pumped upconversion single-photon detector at 1550 nm: performance and noise analysis. *Optics Express* **19,** 21445–21456 (2011).

15. Vandevender, A. P. & Kwiat, P. G. High efficiency single photon detection via frequency up-conversion. *Journal of Modern Optics* **51,** 1433– 1445 (2004).

16. Clark, A. S., Shahnia, S., Collins, M. J., Xiong, C. & Eggleton, B. J. High- efficiency frequency conversion in the single-photon regime. *Optics Letters* **38,** 947–949 (2013).

17. Wright, L. G., Renninger, W. H., Christodoulides, D. N. & Wise, F. W. Nonlinear multimode photonics: nonlinear optics with many degrees of freedom. *Optica* **9,** 824–841 (2022).
12

**Acknowledgements**

A.R.D acknowledges partial support from AFOSR (Grant No FA9550-22-1-0036:P00001), DARPA (Grant No HR00112320021), NASA (Grant No 80NSSC21K095). H.H. acknowledges support from the Department of Energy (Grant No. DE-SC0020101) and the National Science Foundation (Grant No. CBET-2231857). P.V. acknowledges support under the Professional Research Experience Program (PREP), funded by the National Institute of Standards and Technology and administered through the Department of Chemistry and Biochemistry, University of Maryland.

**Author contributions**

P.P.V., H.H., and A.R.D conceived the idea. H.L. fabricated the samples for measurement and performed full-tensor numerical simulations. Y.T. and X.T. performed the linear and nonlinear measurements. H.L. and M.H. performed characterization of fabricated samples using SEM. H.L., P.S., Y.T., and X.T. analyzed the data from linear and nonlinear measurements. H.L., P.S., Y.T., X.T., H.H., and A.R.D. wrote the article with input from all the authors.